\documentclass[twocolumn,superscriptaddress,10pt,floatfix,prb,aps]{revtex4-2}
\ifdefined\pdfminorversion
\fi
\usepackage{amsfonts}
\usepackage{amssymb}
\usepackage{amsmath}
\usepackage{graphicx}
\usepackage{dcolumn}
\usepackage{bm}
\usepackage{units}
\usepackage{multirow}
\usepackage{CJKutf8}
\usepackage{subfigure}
\usepackage{color}%
\usepackage{url}
\usepackage[colorlinks,linkcolor=blue,anchorcolor=blue,citecolor=blue,urlcolor=blue]{hyperref}
\usepackage{microtype}
\makeatletter
\AtBeginDocument{%
  \@booleantrue\raggedcolumn@sw
  \raggedbottom
  \setlength{\parskip}{0pt}%
  \setlength{\textfloatsep}{10pt plus 1pt minus 2pt}%
  \setlength{\dbltextfloatsep}{10pt plus 1pt minus 2pt}%
  \setlength{\floatsep}{7pt plus 1pt minus 1pt}%
  \setlength{\dblfloatsep}{7pt plus 1pt minus 1pt}%
  \setlength{\intextsep}{8pt plus 1pt minus 1pt}%
  \setlength{\abovecaptionskip}{6pt}%
  \setlength{\belowcaptionskip}{1pt}%
  \renewcommand{\section}{\@startsection{section}{1}{\z@}%
    {2ex plus .3ex minus .2ex}{.9ex}%
    {\normalfont\small\bfseries\centering}}%
  \renewcommand{\subsection}{\@startsection{subsection}{2}{\z@}%
    {1.6ex plus .2ex minus .2ex}{.7ex}%
    {\normalfont\small\bfseries\centering}}%
}
\makeatother

\begin{document}

\title{Experimental realization of loss-induced quadrupole topology from a gapped trivial insulator in RLC circuits}

\author{Wenxue He}
\affiliation{Center for Joint Quantum Studies and Tianjin Key Laboratory of Low Dimensional Materials Physics and Preparing Technology, Department of Physics, School of Science, Tianjin University, Tianjin 300350, China}
\author{Chen Wang}
\email[Contact author: ]{physcwang@tju.edu.cn}
\affiliation{Center for Joint Quantum Studies and Tianjin Key Laboratory of Low Dimensional Materials Physics and Preparing Technology, Department of Physics, School of Science, Tianjin University, Tianjin 300350, China}
\author{Fan Yang}
\affiliation{Center for Joint Quantum Studies and Tianjin Key Laboratory of Low Dimensional Materials Physics and Preparing Technology, Department of Physics, School of Science, Tianjin University, Tianjin 300350, China}
\author{Hechen Ren}
\email[Contact author: ]{ren@tju.edu.cn}
\affiliation{Center for Joint Quantum Studies and Tianjin Key Laboratory of Low Dimensional Materials Physics and Preparing Technology, Department of Physics, School of Science, Tianjin University, Tianjin 300350, China}

\date{\today}

\begin{abstract}
Non-Hermitian topological transitions offer a route to controlling topology through dissipation, but their experimental signatures depend on how loss modifies the system's response. Here, we experimentally realize loss-induced quadrupole topology in a square-octagon RLC circuit whose ideal Hermitian limit is a gapped trivial insulator. Patterned grounding resistors introduce on-site loss while the conservative couplings remain fixed. Band spectra and biorthogonal nested Wilson loops identify a loss-driven topological transition accompanied by closure and reopening of the real line gap. Impedance spectroscopy and spatial maps reveal corner-localized responses consistent with the predicted quadrupole phase. Comparing two finite-loss configurations, we observe enhanced corner contrast at stronger patterned loss within the same predicted phase. These results connect dissipation-driven topology to observable boundary responses in RLC circuits.
\end{abstract}
\maketitle

\emph{Introduction.}---Topological band theory conventionally relates bulk topology to boundary states within a Hermitian framework~\cite{hasan_rmp_2010,qi_rmp_2011}, where topology is typically engineered through conservative couplings and symmetries~\cite{su_prl_1979,haldane_prl_1988,kane_mele_prl_2005,clkane_prl_2005,bernevig_science_2006,konig_science_2007,chang_science_2013}, an approach also adopted in classical-wave systems~\cite{haldane_raghu_prl_2008,wu_hu_prl_2015,lu_nphoton_2014,ozawa_rmp_2019,xue_nrm_2022}. Higher-order phases extend this correspondence to corner and hinge states~\cite{benalcazar_science_2017,benalcazar_prb_2017,langbehn_prl_2017,song_prl_2017,ezawa_prl_2018,chen_prl_2020,schindler_sciadv_2018,xie_nrp_2021}. These phases have been realized in mechanical~\cite{serra_garcia_nature_2018}, electromagnetic~\cite{peterson_nature_2018,noh_nphoton_2018,xie_prl_2019}, acoustic~\cite{xue_nmat_2019,ni_nmat_2019,xue_nc_2020}, and electrical platforms~\cite{Imhof_np_2018}, with applications in resonators~\cite{ota_optica_2019,xie_lpr_2020}, wave transport~\cite{wei_prl_2021}, and lasing~\cite{zhang_lsa_2020,kim_nc_2020}. In such designs, dissipation is usually treated as a source of attenuation and resonance broadening rather than a control parameter for topology.
\par

\begin{figure}[!t]
\centering
\includegraphics[width=\linewidth]{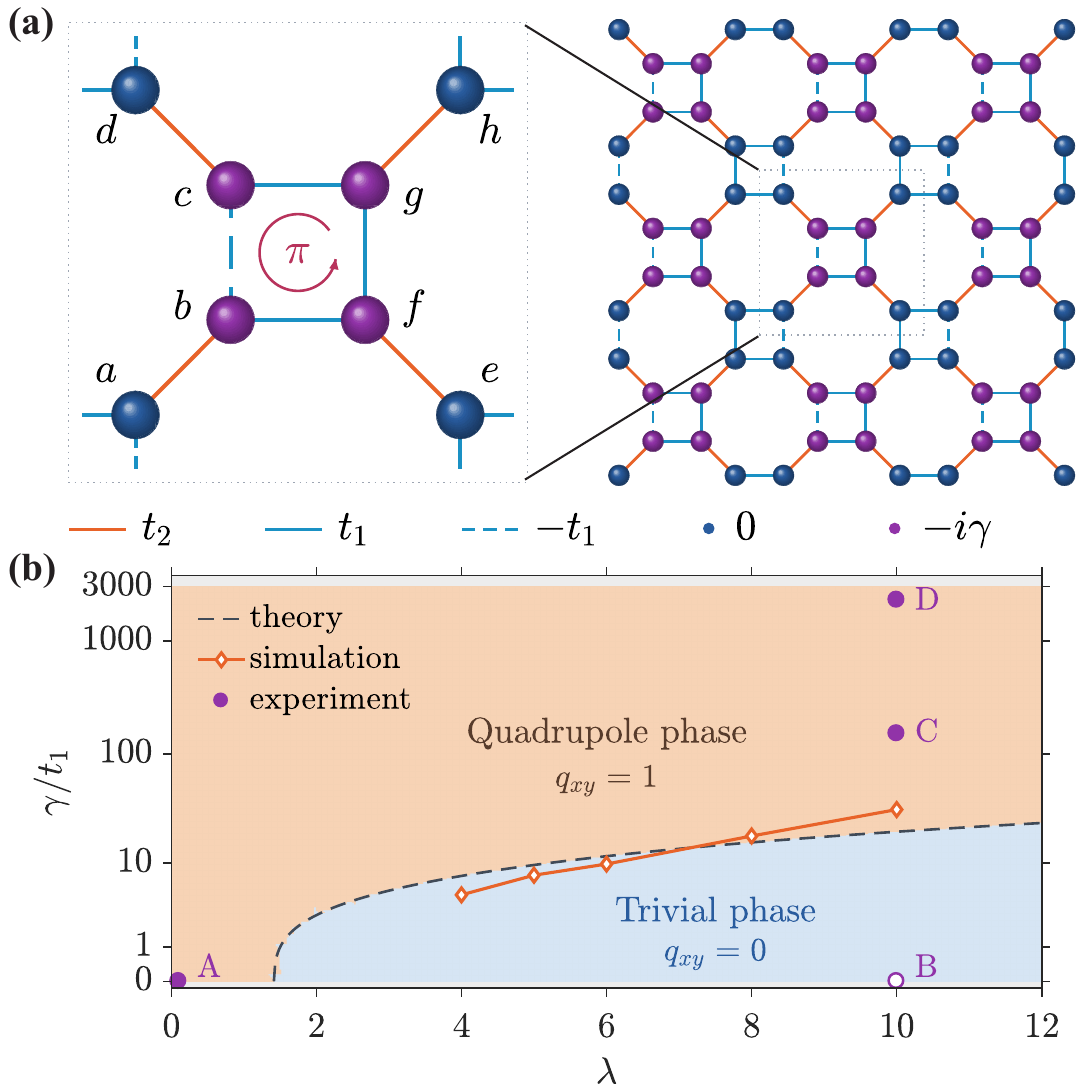}
\caption{\textbf{Square-octagon lattice and phase diagram.} (a) Eight-site square-octagon lattice with a $\pi$ flux through each elementary square. Orange, solid blue, and dashed blue bonds denote $t_2$, $t_1$, and $-t_1$, respectively; purple sites carry on-site loss $-i\gamma$. Taking $t_1>0$ as the hopping scale, we define the dimensionless hopping ratio $\lambda=t_2/t_1$ and loss strength $\bar\gamma=\gamma/t_1$. (b) Quadrupole index $q_{xy}$ versus $\lambda$ and $\bar\gamma$. The dashed curve and orange diamonds denote the analytical bulk phase boundary and simulated corner-response threshold, respectively. Purple circles A--D mark the experimental configurations in Figs.~\ref{fig4}(a)--(d); filled (open) circles indicate resolved (unresolved) corner enhancement.}
\label{fig1}
\end{figure}

Non-Hermitian band theory generalizes topological classification~\cite{shen_prl_2018,gong_prx_2018,kawabata_prx_2019,ashida_advphys_2020,bergholtz_rmp_2021} and provides generalized formulations of bulk--boundary correspondence~\cite{yao_prl_2018,kunst_prl_2018,yokomizo_prl_2019,okuma_prl_2020}. These developments provide a framework for understanding topological transitions driven by gain or loss and their associated boundary states. Engineered dissipation enables topological measurements~\cite{rudner_prl_2009,zeuner_prl_2015} and mode selection~\cite{poli_nc_2015,slootman_prresearch_2024} and can also induce topological phases~\cite{takata_prl_2018,li_prl_2020,liu_prl_2019,luo_prl_2019,zhang_prl_2019,liu_prapplied_2020}. Experimentally, loss-induced topological states have been observed in both first-order~\cite{wetter_prl_2023,hashemi_nmat_2025} and higher-order systems~\cite{gao_nc_2021,wu_prb_2022}, where dissipation opens a topological gap in an initially gapless Hermitian system~\cite{gao_nc_2021,wu_prb_2022}.
\par

A distinct route uses patterned on-site loss to drive an already gapped trivial insulator into a quadrupole phase through bulk-gap closure and reopening at fixed conservative couplings~\cite{tian_prb_2023}. Realizing this route experimentally requires resolving corner signatures while loss changes modal damping and spatial profiles. Studies of disorder effects on localization~\cite{cwang_prresearch_2020,cwang_prb_2024_quadrupole} and boundary transport~\cite{cwang_prb_2021,cwang_prb_2022_hermitian,cwang_prb_2022_hinge}, as well as loss-induced delocalization~\cite{tang_nc_2025_delocalization}, highlight this distinction between topology and response.
\par

Topolectrical circuits offer tunable components and spectral and spatial probes~\cite{lee_comphys_2018,Imhof_np_2018,Yang_PhysRevRes_2020,Wang_NatCommun_2020,helbig_prb_2019,sahin_apled_2025}, including access to non-Hermitian localization~\cite{helbig_np_2020,hofmann_prresearch_2020,Zou_NatCommun_2021,Tang_prb_2023}. Capacitors and inductors implement hopping, while grounding resistors independently control on-site loss. The impedance depends on the inverse admittance matrix and its modal weights~\cite{lee_comphys_2018,schomerus_prresearch_2020}, connecting the predicted topology to observable corner responses.
\par

Here, we realize this route in the square-octagon RLC circuit of Fig.~\ref{fig1}(a). Patterned grounding resistors produce corner-localized responses from a gapped trivial Hermitian limit at fixed capacitive and inductive couplings. Spectra and spatial maps reveal different corner contrasts at two loss strengths within the predicted quadrupole phase. Further simulations test persistence under simultaneous capacitance and inductance disorder.
\par

\emph{Model, topology, and circuit design.}---We consider the eight-site square-octagon lattice in Fig.~\ref{fig1}(a), with hopping amplitudes $t_1$ and $t_2$, a $\pi$ flux through each elementary square, and on-site loss on sublattices $b$, $c$, $f$, and $g$. Its Bloch Hamiltonian at crystal momentum $\bm k$ is
\begin{equation}
H(\bm k)=H_0(\bm k)-i\gamma P_{\mathrm{in}},
\label{eq_1}
\end{equation}
where $H_0$ describes the Hermitian hopping network and $P_{\mathrm{in}}$ projects onto the lossy sublattices. The hopping ratio $\lambda$ and normalized loss $\bar\gamma$ determine the topological phase [Fig.~\ref{fig1}(b)]. In this model, the $\pi$-flux pattern underlies the anticommutation of the two reflection operators, a key ingredient of this quadrupole construction~\cite{benalcazar_science_2017,benalcazar_prb_2017}. The flux is fixed in both the trivial and quadrupole regimes and therefore does not, by itself, distinguish them; see Sec.~S1 of the Supplemental Material (SM)~\cite{supplemental}.
\par

We depict the phase diagram of model~\eqref{eq_1} in Fig.~\ref{fig1}(b) characterized by biorthogonal nested Wilson loops~\cite{benalcazar_prb_2017,luo_prl_2019,tian_prb_2023} and constructed from the four-band subspace whose Bloch eigenvalues $\epsilon_n$ satisfy $\operatorname{Re}\epsilon_n<0$. This subspace is separated from the remaining bands by a real line gap at $\operatorname{Re}\epsilon=0$. With the protecting symmetries and the required band and Wannier-sector gaps intact, the nested polarizations $p_x,p_y$ of the selected Wannier sectors are quantized to $0$ or $1/2$ modulo one. The normalized phase index $q_{xy}=4p_xp_y$ distinguishes the quadrupole ($q_{xy}=1$) and trivial ($q_{xy}=0$) regimes considered here; see numerical details in the SM~\cite[Sec.~S2]{supplemental}.
\par

In the Hermitian limit, where $\bar\gamma=0$, the quadrupole and trivial regimes lie below and above $\lambda_c=\sqrt{2}$, respectively. For fixed $\lambda>\sqrt{2}$, patterned loss drives the gapped trivial insulator into the quadrupole phase at $\bar\gamma_c=2\sqrt{\lambda^2-2}$. As $\bar\gamma$ crosses $\bar\gamma_c$, the bulk real line gap at $\operatorname{Re}\epsilon=0$ closes at the $\Gamma$ point ($\bm k=0$) and reopens in the quadrupole regime. This closing and reopening accompanies a change in $q_{xy}$ from $0$ to $1$. At the experimental coupling ratio $\lambda=10$, the threshold is $\bar\gamma_c\simeq19.8$. The supporting spectra and bulk phase boundary are given in the SM~\cite[Sec.~S3]{supplemental}.
\par

\begin{figure*}[!t]
\centering
\includegraphics[width=0.9\textwidth]{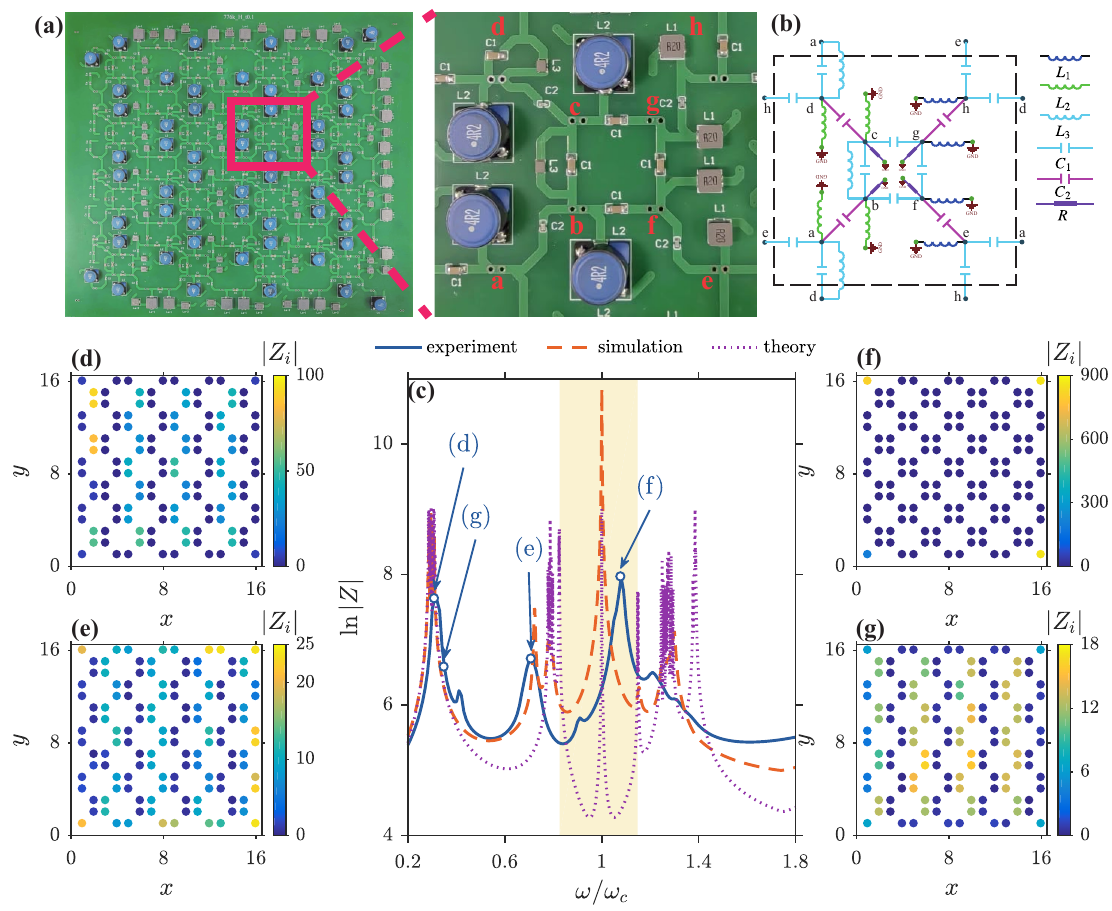}
\caption{\textbf{Hermitian quadrupole topology and impedance responses in LC circuits.} (a) Photograph of the fabricated circuit board and an enlarged view identifying the eight sublattices \(a,b,\ldots,h\). (b) Circuit schematic showing coupling capacitors \(C_1\) and \(C_2\), grounding inductors \(L_1\) and \(L_2\), and parallel \(C_1\)-\(L_3\) branches that implement effective negative couplings at the design frequency. Optional grounding resistors \(R\) are connected to sublattices \(b\), \(c\), \(f\), and \(g\) in the RLC configuration and disconnected in the LC reference shown in (a). (c) Aggregate impedance spectra on a natural-logarithmic scale, comparing experiment (blue), LTspice simulation (orange), and theory (purple). Shading marks the finite-circuit corner-mode frequency window defined in the SM~\cite[Sec.~S5]{supplemental}. (d)--(g) Measured local impedance magnitudes at the frequencies indicated by the corresponding parenthesized labels in (c): (d) \(f=0.241~\mathrm{MHz}\), \( \omega/\omega_c=0.311 \); (e) \(f=0.552~\mathrm{MHz}\), \(\omega/\omega_c=0.710\); (f) \(f=0.839~\mathrm{MHz}\), \(\omega/\omega_c=1.080\); and (g) \(f=0.272~\mathrm{MHz}\), \(\omega/\omega_c=0.350\). The maps show edge-dominated responses in (d) and (e), a corner-dominated response in (f), and a bulk-dominated response in (g). Colors represent \(|Z_i|\) in ohms, with independent color scales for the four maps. The sample contains \(4\times4\) unit cells with open boundaries and \(\lambda=0.1\).}
\label{fig2}
\end{figure*}

To implement this loss-driven topological route, we realize the model in Eq.~\eqref{eq_1} using a capacitor-inductor network with optional resistive grounding~\cite{lee_comphys_2018,Imhof_np_2018,Yang_PhysRevRes_2020}; the circuit is shown in Figs.~\ref{fig2}(a) and (b). Capacitors $C_1$ and $C_2$ set the hopping ratio $\lambda=C_2/C_1$. At the design frequency $\omega_c=[(2C_1+C_2)L_1]^{-1/2}$, parallel $C_1$-$L_3$ branches implement the negative couplings required by the $\pi$-flux pattern, while grounding inductors $L_1$ and $L_2$ provide reactive on-site compensation. Here $\omega$ is angular frequency, with $f=\omega/(2\pi)$ and $f_c=\omega_c/(2\pi)$.
\par

The circuit Laplacian $J(\omega)$ relates the injected-current vector $\bm I$ to the node-voltage vector $\bm V$ through $\bm I=J(\omega)\bm V$. With the convention $e^{i\omega t}$, ideal capacitors, inductors, and resistors have admittances $i\omega C$, $1/(i\omega L)$, and $1/R$, respectively. With the matching conditions in the SM~\cite[Sec.~S4]{supplemental},
\begin{equation}
U^\dagger J(\omega_c)U=i\omega_c C_1\,H/t_1,
\label{eq_2}
\end{equation}
where $H$ is the real-space Hamiltonian and $U$ is a diagonal sublattice gauge transformation that leaves local impedances unchanged. This mapping holds at $\omega_c$; frequency scans use the full $J(\omega)$. 
\par

Grounding sublattices $b$, $c$, $f$, and $g$ through equal resistances $R$ adds $J_R=P_{\mathrm{in}}/R$ and realizes the normalized loss $\bar\gamma=1/(\omega_cRC_1)$. Disconnecting them recovers the ideal Hermitian limit. Thus the capacitances and grounding resistances independently tune the hopping ratio and loss.
\par

For current injected at node $i$ with no external current at the other nodes, the local node-to-ground impedance is $Z_i=V_i/I_i=[J^{-1}]_{ii}$. For an invertible, diagonalizable Laplacian,
\begin{equation}
Z_i(\omega)=\sum_n\frac{\psi_n^{\mathrm R}(i,\omega)[\psi_n^{\mathrm L}(i,\omega)]^*}{j_n(\omega)},
\label{eq_3}
\end{equation}
where $j_n$ are the admittance eigenvalues and $\psi_n^{\mathrm R}$ and $\psi_n^{\mathrm L}$ are the corresponding right and left eigenvectors, normalized by $\langle\psi_m^{\mathrm L}|\psi_n^{\mathrm R}\rangle=\delta_{mn}$. Both the eigenvalues and modal weights therefore determine the measured response. In the ideal lossless LC limit, $J$ is anti-Hermitian at real frequencies and the numerator reduces to $|\psi_n(i,\omega)|^2$; component losses regularize the resonances.
\par

For both LC and RLC implementations, we compare the aggregate impedance $Z(\omega)=\sum_i|Z_i(\omega)|$ with spatial maps of the local node-to-ground impedance, measured by probing each node in turn. The spectra quantify the overall response strength, while the maps reveal its spatial distribution; these maps do not represent a voltage field under a single fixed drive. Spectra are plotted as $\ln[Z/(1\,\Omega)]$, abbreviated as $\ln Z$ throughout. The theoretical spectra include a uniform numerical conductance $\eta\omega_c C_1$ with $\eta=10^{-3}$, which regularizes the ideal-circuit resonances and is not fitted to the measured linewidths.
\par

\emph{Hermitian quadrupole topology.}---We first establish the corner response of the LC configuration without intentional resistive grounding as a reference for the loss-induced experiment. In the ideal-circuit description, this configuration realizes the Hermitian limit of Eq.~\eqref{eq_1}; the measured sample retains residual component losses.
\par

Figure~\ref{fig2}(a) shows the fabricated board and an enlarged view of its eight-site unit cell. The sample contains $4\times4$ unit cells with $C_1=100~\mathrm{nF}$, $C_2=10~\mathrm{nF}$, and $L_1=200~\mathrm{nH}$. These values yield $\lambda=0.1<\lambda_c$ and a nominal design frequency $f_c\simeq0.777~\mathrm{MHz}$. The ideal model therefore lies in the quadrupole phase. 
\par

\begin{figure*}[!t]
\centering
\includegraphics[width=0.9\textwidth]{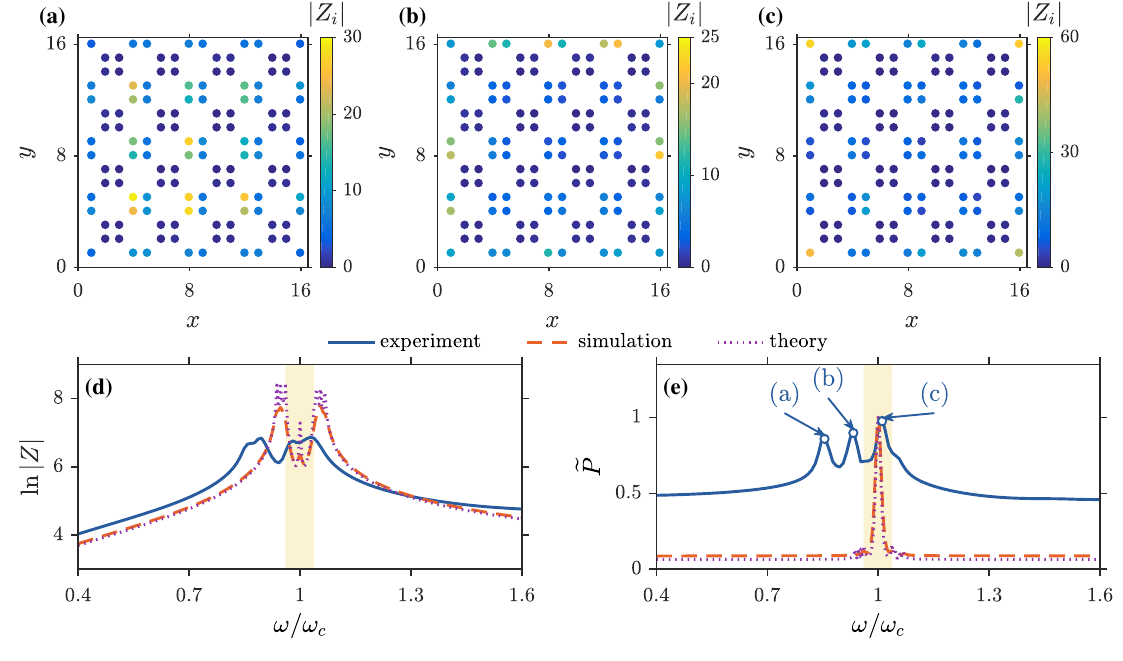}
\caption{\textbf{Loss-induced quadrupole topology and impedance responses in an RLC circuit.} The sample contains \(4\times4\) unit cells with open boundaries, \(\lambda=10\), and patterned grounding resistors with measured mean resistance \(R=6.61~\mathrm{m}\Omega\). (a)--(c) Measured local impedance magnitudes at \(f=0.88\), \(0.96\), and \(1.04~\mathrm{MHz}\), showing predominantly bulk, edge, and corner responses, respectively. Colors indicate \(|Z_i|\) in ohms, with an independent scale for each map. (d) Aggregate impedance spectra plotted as \(\ln Z\), without vertical offsets. (e) Experiment, simulation, and theory curves, each normalized as $\widetilde P=P/P_{\max}$, where $P$ is the response concentration defined in Eq.~\eqref{eq_4} and $P_{\max}$ is that curve's maximum over $0.4\leq\omega/\omega_c\leq1.6$. Normalization compares frequency-dependent features, not absolute concentration. Blue, orange, and purple denote experiment, LTspice simulation, and theory, respectively, in both (d) and (e). The dimensionless frequency axes show \(\omega/\omega_c\), with \(f_c=1.027~\mathrm{MHz}\). Shading marks the theoretical corner-mode frequency window, \(0.961<\omega/\omega_c<1.038\), defined in the SM~\cite[Sec.~S5]{supplemental}.}
\label{fig3}
\end{figure*}

The boards are fabricated from Altium Designer layouts using capacitors and inductors whose measured values are within $4\%$ of their nominal values. Node-to-ground impedances are measured with a TH2826A precision LCR meter after open- and short-circuit calibration. The measurements are compared with full frequency-dependent circuit-Laplacian calculations and LTspice simulations. In the LTspice model, each capacitor and inductor includes a series resistance $R_{\mathrm{ser}}=10~\mathrm{m}\Omega$, distinct from the patterned grounding resistance $R$. Measurement and processing procedures are given in the SM~\cite[Sec.~S5]{supplemental}, and component nonidealities are discussed in the SM~\cite[Sec.~S6]{supplemental}.
\par

Figure~\ref{fig2}(c) shows a sharp theoretical resonance near \(\omega/\omega_c=1\) and a broader measured feature at a higher frequency. At \(\omega/\omega_c=1.080\), the map in Fig.~\ref{fig2}(f) is concentrated at the corners with unequal amplitudes, consistent with the predicted Hermitian corner modes.
\par

The maps at \(\omega/\omega_c=0.311\) and \(0.710\) are edge-dominated [Figs.~\ref{fig2}(d) and \ref{fig2}(e)], whereas that at \(\omega/\omega_c=0.350\) is bulk-dominated [Fig.~\ref{fig2}(g)]. The shaded finite-circuit corner-mode window~\cite[Sec.~S5]{supplemental} neither represents a bulk gap nor guarantees corner dominance throughout. Component tolerances and parasitic impedances can shift and broaden the measured resonances~\cite[Sec.~S6]{supplemental}.
\par

\emph{Loss-induced quadrupole topology.}---We next fix $\lambda=10$, for which the ideal Hermitian model is a gapped trivial insulator, and use patterned loss to access the quadrupole regime. The $4\times4$-cell sample has $C_1=10~\mathrm{nF}$, $C_2=100~\mathrm{nF}$, and $L_1=200~\mathrm{nH}$. After characterizing the circuit without intentional resistive grounding, we connect resistors with measured mean resistance $R=6.61~\mathrm{m}\Omega$ to sublattices $b$, $c$, $f$, and $g$, leaving the reactive couplings unchanged. Measurements and simulations follow the same procedures as for the LC sample.
\par

The nominal components give $f_c\simeq1.027~\mathrm{MHz}$ and $\bar\gamma\approx2.3\times10^3$, well above $\bar\gamma_c$. Using $f_{\mathrm{ref}}=1.05~\mathrm{MHz}$ for the markers in Fig.~\ref{fig1}(b) gives the same loss to two significant figures. These are component-based estimates rather than direct loss measurements. Frequency scans use $f_c$; both loss conventions and the critical-resistance estimates are detailed in the SM~\cite[Sec.~S6]{supplemental}.
\par

\begin{figure*}[!t]
\centering
\includegraphics[width=0.9\textwidth]{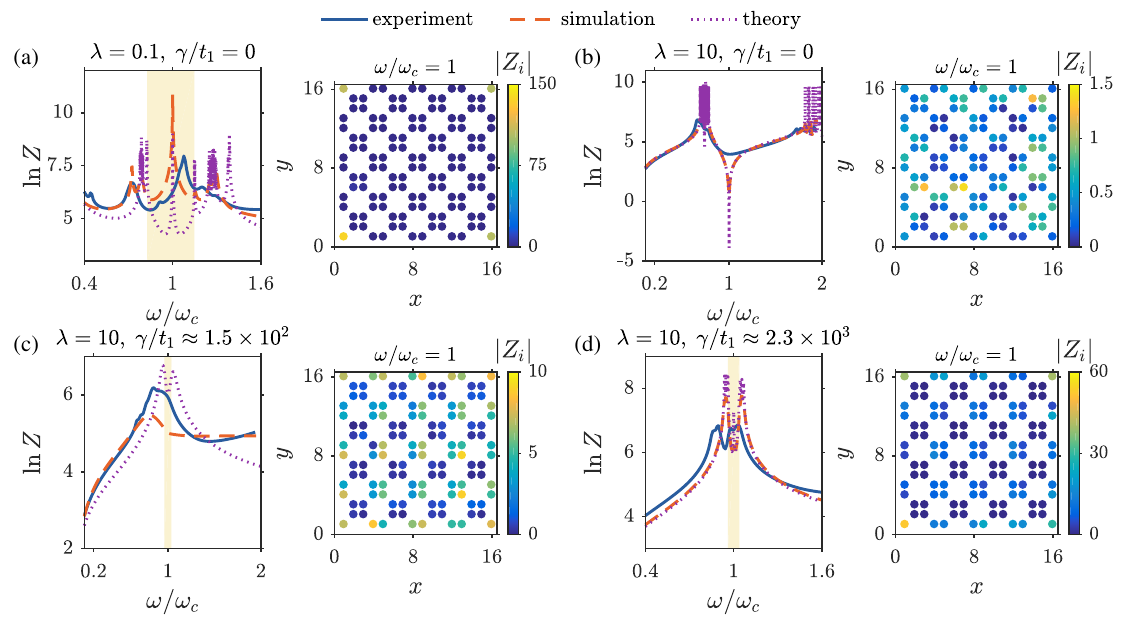}
\caption{\textbf{Control of impedance responses in LC and RLC circuits.} Each panel shows an aggregate impedance spectrum, plotted as $\ln Z$ (left), and a measured local impedance map (right). All maps are measured at the corresponding design frequency $\omega=\omega_c$, marked by $\omega/\omega_c=1$ in each map: $f_c\simeq0.777~\mathrm{MHz}$ in (a) and $1.027~\mathrm{MHz}$ in (b)--(d). Blue, orange, and purple curves denote experiment, simulation, and ideal-circuit theory, respectively. (a) Hermitian quadrupole configuration with \((\lambda,\bar\gamma)=(0.1,0)\). (b) Hermitian trivial configuration with \((\lambda,\bar\gamma)=(10,0)\). (c), (d) \( \lambda=10 \) with nominal loss strengths \(\bar\gamma\simeq1.5\times10^2\) and \( 2.3\times10^3 \), respectively, evaluated at $f_{\mathrm{ref}}$. The grounding resistances are \(R=100~\mathrm{m}\Omega\) in (c) and \(6.61~\mathrm{m}\Omega\) in (d). Both configurations lie in the predicted quadrupole regime but exhibit different corner-response contrasts. Here, \( C_1=100~\mathrm{nF} \) in (a) and \( 10\,\mathrm{nF} \) in (b), (c), and (d), with \( C_2=\lambda C_1 \) and \( L_1=200~\mathrm{nH}\) throughout. Shading marks the finite-circuit corner-mode spectral windows defined in the SM~\cite[Sec.~S5]{supplemental}. Map colors indicate \(|Z_i|\) in ohms, with an independent color scale for each map.}
\label{fig4}
\end{figure*}

Equation~\eqref{eq_3} relates corner visibility to the mode's impedance contribution relative to other modes through its admittance eigenvalue and biorthogonal weight; see the modal analysis in the SM~\cite[Sec.~S6]{supplemental}.
\par

Figures~\ref{fig3}(a)--(c) progress from predominantly bulk to edge to corner responses at the frequencies specified in the caption. The corner enhancement in Fig.~\ref{fig3}(c) is consistent with the predicted loss-induced quadrupole phase. Responses away from the corners can reflect finite localization lengths and other modes, because the impedance probes the circuit Green's function rather than a single eigenstate.
\par

In Fig.~\ref{fig3}(d), the measured aggregate response is broader than the theoretical resonances near \(\omega/\omega_c=1\). Differences can arise from component variations and parasitic impedances, as discussed in the SM~\cite[Sec.~S6]{supplemental}.
\par

To track spatial concentration across the frequency scan independently of the overall impedance amplitude, we define $w_i=|Z_i|/\sum_j|Z_j|$ and obtain
\begin{equation}
P(\omega)=\sum_iw_i^2
=\frac{\sum_i|Z_i(\omega)|^2}{[\sum_i|Z_i(\omega)|]^2}.
\label{eq_4}
\end{equation}
The sums include all $N=128$ nodes, with $1/N\leq P\leq1$ whenever the aggregate impedance is nonzero. Larger values of $P$ indicate stronger spatial concentration. This quantity characterizes the distribution of local impedances measured by sequential node-to-ground probing and differs from the conventional eigenstate inverse participation ratio~\cite{cwang_prb_2024_quadrupole}. The location of the enhanced response must still be determined from the impedance maps.
\par

Figure~\ref{fig3}(e) compares $\widetilde P$ for experiment, simulation, and theory. The original $P$ curves in the SM~\cite[Sec.~S5 and Fig.~S5]{supplemental} differ in absolute magnitude; normalization does not imply quantitative agreement in concentration. The experimental peak near \(\omega/\omega_c=1\) and the map in Fig.~\ref{fig3}(c) jointly identify a localized corner response. Shading denotes the independently calculated corner-mode window~\cite[Sec.~S5]{supplemental}.
\par

\emph{Robustness against disorder.}---To test the stability of the corner response, we also simulate circuits with capacitances and inductances whose relative deviations from their nominal values are uniformly distributed over $[-W,W]$, where $W$ is the maximum relative deviation, as detailed in the SM~\cite[Sec.~S7]{supplemental}. Previous theory identifies a finite-disorder regime of corner localization in both Hermitian and non-Hermitian quadrupole insulators~\cite{cwang_prb_2024_quadrupole}. The selected Hermitian realizations retain pronounced corner concentration, including at $W=20\%$ [SM, Fig.~S3]. In the loss-induced circuit simulated with $R=10~\mathrm{m}\Omega$, corner enhancement also persists: at $W=10\%$, the mean impedance magnitude at the four corner nodes is $12.007$ times the mean over all noncorner nodes [SM, Fig.~S4]. At the larger tested disorder strengths, substantial edge and bulk responses coexist with the corner signal, reducing its contrast relative to the clean circuit. These results, based on a single selected realization at each relative disorder strength $W$, support the persistence of corner enhancement under finite disorder; they do not establish a statistical disorder threshold.
\par

\emph{Tuning corner responses.}---Having established corner localization in the Hermitian and loss-induced configurations, we compare four representative settings in Fig.~\ref{fig4} to examine how the coupling ratio and patterned loss affect the corner response. Corner enhancement is quantified by the ratio of the mean impedance magnitude at the four corner nodes to the mean over all noncorner nodes, including edge nodes. The simulated corner-response threshold in Fig.~\ref{fig1}(b) uses a ratio of at least \(1.1\) at \(\omega=\omega_c\), as specified in the SM~\cite[Sec.~S5]{supplemental}. This operational criterion characterizes response visibility, not the bulk topological invariant.
\par

Without patterned loss, the configuration with \(\lambda=0.1\) exhibits a pronounced corner-localized response near the design frequency; see Fig.~\ref{fig4}(a). Increasing the coupling ratio to \( \lambda=10 \) removes this corner concentration, while resonances associated with other modes remain, as shown in Fig.~\ref{fig4}(b). These observations are consistent with the Hermitian quadrupole and trivial phases, respectively, separated by the critical hopping ratio $\lambda_c$.
\par

At fixed \(\lambda=10\), Figs.~\ref{fig4}(c) and (d) compare nominal losses \(\bar\gamma\simeq1.5\times10^2\) and \(2.3\times10^3\), evaluated at $f_{\mathrm{ref}}$. Both lie above $\bar\gamma_c$ in the predicted quadrupole regime. Substantial noncorner responses limit the contrast in (c), whereas (d) shows stronger corner concentration: the corner-to-noncorner ratios at the respective design frequencies $\omega=\omega_c$ are \(2.677\) and \(7.671\)~\cite[Sec.~S5]{supplemental}. Comparing only with noncorner nodes without intentional grounding loss gives \(1.452\) and \(3.870\), respectively~\cite[Sec.~S5]{supplemental}, confirming increased visibility within the same predicted phase. For (a), (b), and (d), the measured spectra qualitatively agree with theory and simulations, while the maps support the predicted presence or absence of corner localization.
\par

The contrast difference between Figs.~\ref{fig4}(c) and (d) is consistent with Eq.~\eqref{eq_3}, although the measurements do not isolate a unique cause. In the strong-loss limit, eliminating the resistively grounded sublattices yields induced damping proportional to $\lambda^2/\bar\gamma$ in the retained subspace~\cite[Sec.~S6]{supplemental}. Increasing loss can reduce this damping and help resolve corner contributions. Because the leading correction is common to the retained subspace, contrast also depends on noncorner modes and residual circuit losses. This offers a qualitative explanation, not a quantitative fit. Along the present tuning path, stronger loss improves corner visibility within the same predicted phase; a nontrivial bulk index alone does not guarantee a corner-dominated response.
\par

\emph{Conclusion.}---We have experimentally realized loss-induced quadrupole topology in a square-octagon RLC circuit whose ideal Hermitian limit is a gapped trivial insulator. Patterned grounding resistors tune on-site loss while the capacitive and inductive hopping network remains fixed. Impedance spectra and spatial maps reveal corner-localized responses consistent with the predicted quadrupole phase, with markedly different degrees of corner contrast at the two loss strengths. Disorder simulations further show that corner enhancement survives finite variations of both capacitances and inductances in the tested realizations. These observations demonstrate loss-controlled higher-order boundary responses and show that their visibility depends on modal contributions as well as the bulk topological phase.
\par

\begin{acknowledgments}

This work is supported by the Fundamental and Interdisciplinary Disciplines Breakthrough Plan of the Ministry of Education of China (JYB2025XDXM410) and the National Natural Science Foundation of China under Grant No.~12574023. 
\par

\emph{Data availability.}---The raw data and the data-analysis code supporting the findings of this Letter are openly available in Code Ocean at \href{https://codeocean.com/capsule/4765365/tree/v1}{https://doi.org/10.24433/CO.6211221.v1}.\par

\end{acknowledgments}

\bibliographystyle{apsrev4-2}
\bibliography{references}

\end{document}